\documentclass{elsart}

\usepackage{epsfig}

\usepackage{amssymb}

\newcommand{\continueend}[0]{\raisebox{4mm}%
{\rotatebox{-90}{$\curvearrowright$}}}
\newcommand{\continuebeg}[0]{\raisebox{-1mm}%
{\rotatebox{90}{$\curvearrowleft$}}}

\begin{document}

\begin{frontmatter}

\begin{flushright}
hep-ph/0609017\\
CERN-LCGAPP-2006-03\\
September 2006\\[8mm]
\end{flushright}

\title{A standard format for\\ Les Houches Event Files}

\author[louvain]{J. Alwall}
\author[torino]{A. Ballestrero}
\author[gainesville]{P. Bartalini}
\author[dubna]{S. Belov}
\author[moscow]{E. Boos}
\author[durham]{A. Buckley}
\author[london]{J.M. Butterworth}
\author[moscow]{L. Dudko}
\author[cern,genova]{S. Frixione}
\author[fnal]{L. Garren}
\author[karlsruhe]{S. Gieseke}
\author[protvino]{A. Gusev} 
\author[berkeley]{I. Hinchliffe}
\author[lansing]{J. Huston}
\author[ljubljana]{B. Kersevan}
\author[dresden]{F. Krauss}
\author[lund]{N. Lavesson}
\author[lund]{L. L\"onnblad}
\author[torino]{E. Maina}
\author[louvain]{F. Maltoni}
\author[cern]{M.L. Mangano}
\author[zurich]{F. Moortgat}
\author[fnal]{S. Mrenna}
\author[athens]{C.G. Papadopoulos}
\author[torino]{R. Pittau}
\author[durham]{P. Richardson}
\author[cern,manchester]{M.H. Seymour}
\author[cambridge]{A. Sherstnev}
\author[cern,lund]{T. Sj\"ostrand\corauthref{corts}}
\author[fnal]{P. Skands}
\author[protvino]{S.R. Slabospitsky} 
\author[cracow]{Z. W\c{a}s}
\author[cambridge]{B.R. Webber} 
\author[cracow]{M. Worek}
\author[karlsruhe]{D. Zeppenfeld}

\address[louvain]{Centre for Particle Physics and Phenomenology (CP3),
      Universit\'{e} Catholique de Louvain, Chemin du Cyclotron 2,
      B--1348 Louvain-la-Neuve, Belgium}
\address[dresden]{Institute for Theoretical Physics, TU Dresden,
      D--01062 Dresden, Germany}
\address[karlsruhe]{Institut f\"ur Theoretische Physik, 
      Universit\"at Karlsruhe, D--76128 Karlsruhe, Germany}
\address[athens]{Institute of Nuclear Physics, NCSR ``Demokritos''
      15310 Athens, Greece}
\address[genova]{INFN, Sezione di Genova, Via Dodecaneso 33, 
      I--16146 Genova, Italy}
\address[torino]{INFN, Sezione di Torino and Dip. Fisica Teorica, 
      Universit\'a di Torino, via Giuria 1, I--10125 Torino, Italy}
\address[cracow]{Institute of Nuclear Physics, PAN, 
      ul. Radzikowskiego 152, 31-3420 Krak\'ow, Poland} 
\address[dubna]{Joint Institute for Nuclear Research, Joliot-Curie 6,
      Dubna, 141980 Moscow region, Russia} 
\address[moscow]{Skobeltsyn Institute of Nuclear Physics, MSU,
      119992 Moscow, Russia}
\address[protvino]{Institute for High Energy Physics, Protvino, 
      Moscow Region, Russia}
\address[ljubljana]{Faculty of Mathematics and Physics, University of 
      Ljubljana, Jadranska 19, SI-1000 Ljubljana, Slovenia}
\address[lund]{Department of Theoretical Physics, Lund University,
      S\"olvegatan 14A, SE--223 62 Lund, Sweden}
\address[cern]{CERN/PH, CH--1211 Geneva 23, Switzerland}
\address[zurich]{Institute for Particle Physics, ETH Zurich,
      CH--8093 Zurich, Switzerland} 
\address[cambridge]{Cavendish Laboratory,University of Cambridge, 
      J J Thomson Avenue, Cambridge CB3 0HE, United Kingdom} 
\address[durham]{Institute for Particle Physics Phenomenology,
      University of Durham, South Rd, Durham DH1 3LE, United Kingdom}
\address[london]{Physics and Astronomy Department, University College 
      London, Gower St, London WC1E 6BT, United Kingdom}
\address[manchester]{School of Physics and Astronomy, 
      The University of Manchester, Manchester, M13 9PL, 
      United Kingdom}
\address[fnal]{FNAL, P.O. Box 500, Batavia, IL 60510, USA}
\address[berkeley]{Lawrence Berkeley National Lab, Berkeley, CA 94720, 
      USA} 
\address[lansing]{Michigan State University, East Lansing, MI 48840,
      USA}
\address[gainesville]{Department of Physics, University of Florida, 
      Gainesville, FL 32611, USA}

\corauth[corts]{Corresponding author. 
\textit{Email address}: \texttt{torbjorn@thep.lu.se}.}

\begin{abstract}
A standard file format is proposed to store process and event
information, primarily output from parton-level event generators 
for further use by general-purpose ones. The information content is 
identical with what was already defined by the Les Houches Accord
five years ago, but then in terms of Fortran commonblocks. This 
information is embedded in a minimal XML-style structure, for 
clarity and to simplify parsing. 
\end{abstract}

\begin{keyword}
event generators \sep Les Houches Accord \sep file format
\PACS 07.05.Kf \sep 13.85.-t \sep 13.87.-a
\end{keyword}

\end{frontmatter}

\section{Introduction}

The (original) Les Houches Accord (LHA) for user-defined processes 
\cite{lha} has been immensely successful. It is routinely used to 
pass information from matrix-element-based generators to 
general-purpose ones, in order to generate complete events for 
a multitude of processes. The original standard was in terms of two 
\textit{Fortran commonblocks} where information could be stored, 
while the actual usage has tended to be mainly in terms of 
\textit{files} with parton-level events, and increasingly will be 
used by \textit{C++} generators. 
Since the format of such event files is not specified by the 
standard, several different formats are in current usage. This 
leads to a duplication of effort when such files are to be parsed,
a problem that may increase when more programs are developed for
different LHC physics aspects. 

Following the recent discussions at the MC4LHC-06 workshop at CERN,
and subsequent e-mail exchanges, an agreement on a Les Houches Event 
File (LHEF) format has been reached among the signing authors,
representing several of the most commonly-used parton-level and
general-purpose generators, as well as interested end-users.
This standard should allow all information specified
by the LHA to be read in from any file that is LHEF-compliant, and
read using one common parser. 

The current agreement only standardizes the storage of information 
already defined by the LHA, and its wrapping in a lightweight XML-style
structure. In order to allow further information to be stored,
however, comment lines can be appended to the compulsory information.

\section{Existing commonblocks}

In the LHA two commonblocks are used to store data. A brief summary
follows, as a reminder, but anyone not familiar with the details is 
referred to the original publication \cite{lha}.

Initialization information is stored in \texttt{HEPRUP}: 
\begin{verbatim}
      INTEGER MAXPUP
      PARAMETER (MAXPUP=100)
      INTEGER IDBMUP,PDFGUP,PDFSUP,IDWTUP,NPRUP,LPRUP
      DOUBLE PRECISION EBMUP,XSECUP,XERRUP,XMAXUP
      COMMON/HEPRUP/IDBMUP(2),EBMUP(2),PDFGUP(2),PDFSUP(2),
     &IDWTUP,NPRUP,XSECUP(MAXPUP),XERRUP(MAXPUP),
     &XMAXUP(MAXPUP),LPRUP(MAXPUP)
\end{verbatim}
The first few variables refer to the two incoming beams: identities
(\texttt{IDBMUP}), energies (\texttt{EBMUP}), and PDF sets used
(\texttt{PDFGUP, PDFSUP}). \texttt{IDWTUP} defines the weighting
strategy to be used; e.g., 3 corresponds to accepting all events
as they come. The rest defines a set of \texttt{NPRUP} 
separately identified processes, with cross-section information 
(\texttt{XSECUP, XERRUP, XMAXUP}) and an integer label (\texttt{LPRUP})
for each.  

Information on each separate event is stored in \texttt{HEPEUP}:
\begin{verbatim}
      INTEGER MAXNUP
      PARAMETER (MAXNUP=500)
      INTEGER NUP,IDPRUP,IDUP,ISTUP,MOTHUP,ICOLUP
      DOUBLE PRECISION XWGTUP,SCALUP,AQEDUP,AQCDUP,PUP,VTIMUP,
     &SPINUP
      COMMON/HEPEUP/NUP,IDPRUP,XWGTUP,SCALUP,AQEDUP,AQCDUP,
     &IDUP(MAXNUP),ISTUP(MAXNUP),MOTHUP(2,MAXNUP),
     &ICOLUP(2,MAXNUP),PUP(5,MAXNUP),VTIMUP(MAXNUP),
     &SPINUP(MAXNUP)  
\end{verbatim}
Here \texttt{NUP} is the number of particles in the event, with
each particle characterized by its identity (\texttt{IDUP}, using 
the standard PDG numbering \cite{rpp}), status (\texttt{ISTUP}), 
mother(s) (\texttt{MOTHUP}), colours(s) (\texttt{ICOLUP}), 
four-momentum and mass (\texttt{PUP}), proper lifetime (\texttt{VTIMUP}) 
and spin (\texttt{SPINUP}). In addition the event as a whole is 
characterized by the an event weight (\texttt{XWGTUP}), a scale 
(\texttt{SCALUP}), and the $\alpha_{\mathrm{em}}$ (\texttt{AQEDUP}) 
and $\alpha_{\mathrm{s}}$ (\texttt{AQCDUP}) values used.

Only in one respect is there any need to update the meaning of the
variables: the original PDF set numbering in \texttt{PDFGUP} and 
\texttt{PDFSUP} was based on the then-prevalent PDFLIB library 
\cite{pdflib}. Now usage has shifted towards LHAPDF \cite{lhapdf}, 
where a set is specified by a file name and a number, and its LHAGLUE 
\cite{lhaglue} interface, where these two are mapped onto a single 
number. Users of LHAGLUE are recommended to set \texttt{PDFGUP} equal 
to 0, a value that does not clash with PDFLIB, and set \texttt{PDFSUP} 
equal to the LHAGLUE code. We recognize that other schemes may be 
required in the future, with other \texttt{PDFGUP} values.

\section{Basic write and read}

The basic principle for an LHE file is that, in the representation of 
both commonblocks, an initial line should contain all information fields 
that only appear once, while as many subsequent lines as required follow, 
i.e.\ \texttt{NPRUP} or \texttt{NUP} ones, each with information on one 
process or particle, respectively. On each line, all variables to appear 
there should be written out in the same order as they appear in the 
commonblocks, and \textit{with no omissions}. Each value is separated by 
at least one blank from the preceding one, so that free-format reading is 
possible.

Wrapper and comment lines will be allowed in the file, as outlined in the 
next section. Omitting such lines, the rules above lead to a unique 
structure for a file:\\[2mm] 
1) Initialization information, given once\\
a) one line with process-number-independent information:\\
\hspace*{5mm}\texttt{IDBMUP(1) IDBMUP(2) EBMUP(1) EBMUP(2) PDFGUP(1) %
PDFGUP(2)}~\continueend\\
\hspace*{2mm}\continuebeg~\texttt{PDFSUP(1) PDFSUP(2) IDWTUP NPRUP}\\
b) \texttt{NPRUP} lines, one for each process \texttt{IPR} in the range 
1 through \texttt{NPRUP}:\\
\hspace*{5mm}\texttt{XSECUP(IPR) XERRUP(IPR) XMAXUP(IPR) LPRUP(IPR)}\\[2mm]
2) Event information, repeated as many times as there are events\\
a) one line with common event information:\\
\hspace*{5mm}\texttt{NUP IDPRUP XWGTUP SCALUP AQEDUP AQCDUP}\\
b) \texttt{NUP} lines, one for each particle \texttt{I} in the range
1 through \texttt{NUP}\\
\hspace*{5mm}\texttt{IDUP(I) ISTUP(I) MOTHUP(1,I) MOTHUP(2,I) ICOLUP(1,I)}%
~\continueend\\
\hspace*{2mm}\continuebeg~%
\texttt{ICOLUP(2,I) PUP(1,I) PUP(2,I) PUP(3,I) PUP(4,I) PUP(5,I)}%
~\continueend\\
\hspace*{2mm}\continuebeg~\texttt{VTIMUP(I) SPINUP(I)}\\[2mm]
The continuation arrows $\continueend$ and $\continuebeg$ are there 
to emphasize that what is (has to be!) a single line in a file here 
has been split for reasons of visibility.

We stress that all fields must be written out, to make reading predictable. 
When space is at premium, there is no need to have more than one blank 
between fields, to align output in columns, or to use more precision 
than required. For instance, for particles that do not produce a secondary
vertex (or have no spin information) it is enough to write ``0.'' 
(or ``9.'', the default value for unknown spin) in the \texttt{VTIMUP(I)} 
(or \texttt{SPINUP(I)}) field.

\section{Complete file format}

All the above data has to be stored in one single file, together
with further program-specific information on how the events were 
generated, i.e.\ all input needed to reproduce the event sample
in the file: generator name and version, processes associated with 
the respective \texttt{LPRUP(IPR)} label, masses, couplings, 
cuts, etc., as required. Also further information specific to 
each event can be added, such as the actual parton density values 
or multiple scales needed for matrix-element-to-shower matching. 
Currently, such information has to be generator-specific, since 
it would not be a trivial task to define a common scheme to cover 
every possible case. This further information must be clearly 
distinguishable from the compulsory one already described in the 
previous section, however.

To this end, an XML-like structure has been introduced, where
XML-style tags provide blocks where the compulsory and optional 
information is to be found. The file then looks like 
\begin{verbatim} 
<LesHouchesEvents version="1.0">
  <!--
    # optional information in completely free format,
    # except for the reserved endtag (see next line)
  -->
  <header>
    <!-- individually designed XML tags, in fancy XML style -->
  </header>
  <init>
    compulsory initialization information
    # optional initialization information
  </init>
  <event>                            
    compulsory event information       
    # optional event information       
  </event>                           
  (further <event> ... </event> blocks, one for each event)
</LesHouchesEvents>
\end{verbatim}
Indentation of tags and information is allowed (but not required), 
but the tags defined above must be alone on their respective line. 
This is crucial for the \texttt{<init>} and \texttt{<event>} tags, 
which define the points after which the compulsory information is 
to be found. 

The all-enclosing \texttt{LesHouchesEvents} block defines the root
element required by XML, and makes it obvious which standard
and version the file is based on. 

Information on how events were generated --- whatever the author
deems relevant --- should appear in the beginning of the file, 
inside either or both of the \texttt{<!-- ... -->} and 
\texttt{<header> ... </header>} blocks. Both blocks are optional, 
and the order between them can be reversed. The difference is that 
the former block can be written in any format desired, except for 
the reserved endtag, while the \texttt{header} block has to be based 
on pure XML syntax, so that standard XML parsers could be used to 
extract any specific piece of information. Any comments in this block 
must also be of the standard \texttt{<!--  .... -->} kind.  

The \texttt{init} block contains the \texttt{HEPRUP} initialization 
information, and each \texttt{event} block the \texttt{HEPEUP} 
event information. To ensure simple parsing, the \texttt{<init>} line 
is \textit{immediately} followed by the initialization information as 
strictly defined by point 1 in the previous section. Correspondingly, 
each \texttt{<event>} line is \textit{immediately} followed by the 
event information defined by point 2 above.

The advantage of this scheme is that parsing is made simple, since
the format to use next is always known, at least for the compulsory
information. That is, a character string is read from each line,
repeatedly, until either \texttt{<init>} or \texttt{<event>}
is encountered. Then the next line has to be of the form 1a) or 2a), 
respectively, and the subsequent \texttt{NPRUP} or \texttt{NUP} 
ones of 1b) or 2b).

To allow for the future introduction of attributes in \texttt{<init>} 
and \texttt{<event>}, so long as they are all on one line, 
\texttt{<init} or \texttt{<event} strings followed by a blank (and 
then further text on the same line) should be considered as equivalent 
to the normal forms, in terms of how a file is parsed for Les Houches 
standard information. Also \texttt{<header>} might acquire attributes, 
but that would of course not affect the LHA information parsing.
Private attributes are allowed, but with the understanding that any 
future standard would take precedence in case of conflict.  

Any number of comment lines can follow \textit{after} the compulsory 
initialization or event information. These lines can contain whatever 
information desired in whatever format desired, except for the  
already reserved tags. (Anybody inserting an \texttt{<event>} tag as
part of a comment should expect parsing to derail.) However, to avoid
conflicts with XML parsers, it is recommendable to avoid the 
reserved symbols \texttt{<},\texttt{>} and \texttt{\&}, except 
as part of some proper XML syntax. 

Furthermore, it is recommended (but not required) to begin comment lines 
with an \texttt{\#} symbol, to make them easily distinguishable from the 
compulsory information. As an example, there is a demand from the 
experimental community that the parton densities used event by event 
should be stored somewhere, for use when reweighting to another 
parton-density set. This could be accommodated with an optional line 
of the form\\
\texttt{ \#pdf id1 id2 x1 x2 scalePDF xpdf1 xpdf2}\\
where \texttt{\#pdf} is an identifying label and the rest are the 
respective values (identity \cite{rpp}, $x$, $Q$, $xf_i(x,Q)$).

To read a large file, with many events, is slower with a 
conventional XML parser than with simple Fortran/C++ code, potentially
by orders of magnitude. This is a main reason to keep the event data
format simple. The initial information is only to be read once, 
however, and could therefore use a more sophisticated XML structure. 
It is then assumed that normal XML parsing could be stopped after 
\texttt{</init>}, as is possible with some of the existing parsers,
while ones that load the whole document would be unsuitable.

To summarize, the objective of the current standard is to promote a
structure that is completely consistent with XML, so that XML parsers
could be used if desired, but also simple enough that all key information 
should be easily obtainable with fast Fortran/C++ code. 
It would then not be wise to classify a file as 
useless simply because it happens to contain some minor XML-syntactical 
error, like an \texttt{\&} symbol in a comment. A sensible 
strategy for writing a minimal event-generator parser is only to rely 
on the \texttt{<init}  and \texttt{<event} tags opening their respective 
line, being followed on consecutive lines by the respective compulsory 
information, and consider everything else as potentially 
program-dependent.

\section{Outlook}

Several program authors are committed to implementing the standard
in the near future, so that it should rapidly come to ease the
intercommunication burden. The webpage\\[2mm] 
\hspace*{10mm}\texttt{http://www.thep.lu.se/}%
$\sim$\texttt{torbjorn/lhef}\\[2mm]
contains some early examples of Fortran and C++ implementations of parsers.

To allow easy identification of files that follow the standard we
propose that these be given a \texttt{.lhe} file name extension. 

The current proposal must not be viewed as the end of the road. 
There is much further information exchange that ought to be standardized.
It is allowed to use/promote a ``private standard'' of tags in the 
\texttt{header} block or of additional event information, 
and experience with such could point the way towards an extended 2.0 
standard at a later date. Such complementary work has been performed 
within the MCDB project \cite{mcdb} and will continue within the
LCG/CEDAR HEPML project \cite{hepml}.

\end{document}